\documentclass{tMPH2e}
\usepackage{graphicx}
\usepackage{amsmath}

\newcommand\HF{{\mathrm{HF}}}

\begin{document}

\markboth{Grassi \emph{et al.}}{Molecular dissociation energies and heats of
formation within the BOCE approach}

\title{Equilibrium molecular energies used to obtain\break
molecular dissociation energies and heats of formation\break
within the bond-order correlation approach}

\author{A. Grassi, G. M. Lombardo, G. Forte\\
\affil{Dipartimento di Scienze Chimiche, Facolt\`a di Farmacia, Universit\`a di Catania,\break
Viale A. Doria, 6, I-95126 Catania, Italy}
G. G. N. Angilella,$^\ast$\thanks{$^\ast$ Corresponding author.
E-mail: {\tt giuseppe.angilella@ct.infn.it}.} R. Pucci
\affil{%
Dipartimento di Fisica e Astronomia, Universit\`a di Catania, and\break
Lab. MATIS-INFM, and CNISM, Sez. Catania, and\break
Istituto Nazionale di Fisica Nucleare, Sez. Catania,\break
Via S. Sofia, 64, I-95123 Catania, Italy}
N. H. March
\affil{Abdus Salam International Centre for Theoretical Physics, Trieste, Italy, and\break
Oxford University, Oxford, UK}}

\received{Received: 15 August 2005}

\maketitle

\abstract{%
\emph{Ab initio} calculations including electron correlation are still
extremely costly except for the smallest atoms and molecules. Therefore, our
purpose in the present study is to employ a bond-order correlation approach to
obtain, via equilibrium molecular energies, molecular dissociation energies and
heats of formation for some 20 molecules containing C, H, and O atoms, with a
maximum number of electrons around 40. Finally, basis set choice is shown to be
important in the proposed procedure to include electron correlation effects in
determining thermodynamic properties. With the optimum choice of basis set, the
average percentage error for some 20 molecules is approximately 20\% for heats
of formation. For molecular dissociation energies the average error is much
smaller: $\sim 0.4$\%.
}

\section{Introduction}
\label{sec:intro}

In the last few years, much effort has been devoted to the development of
theoretical techniques for the calculation of fundamental thermodynamical
quantities related to molecular formation. These quantities include the
molecular heat of formation, $\Delta H$, the variation of entropy, $\Delta S$,
and the variation of free energy, $\Delta G$, as well as other physical
quantities related to them, such as the molecular dissociation energy $D_0$,
and the equilibrium constants of the chemical reactions.

The available theoretical techniques to extract such quantities are especially
helpful whenever the experimental results, albeit accurate, are affected by
uncertainties, due to the difficulty of obtaining pure samples of a given
compound. Therefore, efficient and accurate theoretical methods are required to
compute thermodynamical quantities.

In particular, the G$x$ methods ($x=1,2,3$) have been recently developed to
this purpose \cite{Curtiss:91}. These techniques are based on the combination
of several \emph{ab initio} molecular energy calculations, using various basis
sets, and including correlation effects within the framework of
M\o{}ller-Plesset (MP) theory \cite{Moeller:34}, at MP2 and MP4 levels of
approximation. However, one of the main drawbacks of the G$x$ methods is their
computational complexity. Indeed, even for the energy calculation of molecules
of relatively small size, both computation time and memory occupancy are quite
expensive \cite{note:computational}\nocite{Curtiss:98}. As a consequence, the
application of this class of methods is effectively limited by the system size.
Therefore, it is of considerable interest to investigate alternative
theoretical techniques to calculate useful thermodynamical quantities for the
formation of molecules, which require less computational resources, while
maintaining a good level of accuracy.

In this context, Cremer \emph{et al.} \cite{Cremer:82,Cremer:82a} have
demonstrated that it is possible to extract the Schr\"odinger energy of a
molecule, $E_S$, \emph{i.e.} the energy corresponding to the exact solution of
the Schr\"odinger equation for a given molecule, from the experimental values
of some observable quantities, such as the molecular heat of formation and the
molecular vibrational frequencies. Inversely, any theoretical model able to
closely reproduce the molecular Schr\"odinger energies, is also expected to
yield molecular heats of formation $\Delta H$ close to the experimental ones.

By definition, the difference between the molecular Schr\"odinger energy $E_S$
and the molecular energy obtained from a calculation at Hartree-Fock (HF) level
is the `experimental' correlation energy, $E_c^{\exp}$. For this reason, in
Ref.~\cite{Grassi:96} we developed a semi-empirical method to calculate a
theoretical estimate $E_c^{\mathrm{theo}}$  (the bond-order correlation energy,
or BOCE) to the correlation energy $E_c^{\exp}$ based on the molecular
bond-order matrix. Within the framework of the Lewis model \cite{Lewis:24},
which describes the bond between atoms in a molecule as the sum of localized
electron pairs, this `theoretical' correlation energy $E_c^{\mathrm{theo}}$ is
calculated as the sum of the correlation energies of each atom, and the
contribution due to the binding energy of each bond. The latter is in turn
evaluated in terms of the product between the molecular bond-order matrix and
some empirical parameters (see also Ref.~\cite{Grassi:04a} for more details).
Since the BOCE method relies on a relatively simple transformation of the
matrix of the molecular orbital coefficients, it turns out that it is a rather
fast and accurate technique, as compared to the G$x$ methods.

The BOCE technique has been applied to calculate molecular properties, such as
the molecular energy $E_c^{\mathrm{theo}}$, the ionization potentials, and the
electron affinities of closed and open shell molecules containing C, H, and O
atoms \cite{Grassi:96}, and recently it has been extended to closed and open
shell molecules containing Si, H, F, and Cl \cite{Grassi:04a}. In view of the
previous satisfactory results for these molecular classes, in this paper we
apply the BOCE method to the calculation of other molecular properties such as
the molecular heat of formation and the molecular dissociation energies of
several molecules containing C, H, and O. Since the bond-order matrix depends
on the basis set used in the HF calculation, we compare the results within the
G2 model with our BOCE results obtained using different basis sets.

\section{Method}

The starting point of the BOCE method is the experimental correlation energy,
$E_c^{\exp}$. This is defined as the difference between the Schr\"odinger
molecular energy $E_S$ and the Hartree-Fock energy, $E_\HF$,
\begin{equation}
E_c^{\exp} = E_S - E_\HF .
\label{eq:S-HF}
\end{equation}
In Eq.~(\ref{eq:S-HF}), $E_\HF$ is the energy from an ideal Hartree-Fock
calculation, requiring an infinite basis set. It should be emphasized that, in
any practical calculation, $E_\HF$ includes errors due to the particular choice
of basis set in the HF calculation. These errors can be made arbitrarily low,
by choosing a sufficiently large basis set.

In analogy to Eq.~(\ref{eq:S-HF}), one can define a `theoretical' correlation
energy $E_c^{\mathrm{theo}}$, as the difference between the molecular energy
$E^{\mathrm{theo}}$ obtained within a theoretical model at higher level than HF
(below, the superscript `theo' will alternatively refer to G$x$ or BOCE), and 
the HF molecular energy:
\begin{equation}
E_c^{\mathrm{theo}} = E^{\mathrm{theo}} - E_\HF .
\label{eq:t-HF}
\end{equation}
Following the molecular dissociation scheme, reported in Ref.~\cite{Grassi:04a},
\begin{equation}
AB \longrightarrow A + B + \mbox{binding energy},
\end{equation}
both experimental and theoretical correlation energies can be partitioned into
sums of atomic and bond contributions,
\begin{subequations}
\begin{eqnarray}
\label{eq:Eexp}
E_c^{\exp} &=& \sum_A E_c^{\exp} (A)  + \sum_{\mbox{\scriptsize all $AB$}} E_c^{\exp} (AB)
,\\
\label{eq:Etheo}
E_c^{\mathrm{theo}} &=& \sum_A E_c^{\mathrm{theo}} (A)  + \sum_{\mbox{\scriptsize all $AB$}} E_c^{\mathrm{theo}} (AB)
,
\end{eqnarray}
\end{subequations}
where the sums run over all atoms and pairs of atoms in the molecule,
respectively, and both experimental and theoretical atomic correlation energies
are defined as in Eqs.~(\ref{eq:S-HF}) and (\ref{eq:t-HF}), respectively.

As assumed in Refs.~\cite{Grassi:96,Grassi:04a}, within the BOCE method
($\mbox{theo}=\mbox{BOCE}$) we identify the theoretical energy of a single atom with 
its Schr\"odinger energy,
\begin{equation}
E_S (A) = E^{\mathrm{BOCE}} (A).
\label{eq:identify}
\end{equation}
Therefore, the atomic theoretical correlation energies will be equal to the
experimental ones:
\begin{equation}
E_c^{\mathrm{BOCE}} (A) = E_c^{\exp} (A).
\end{equation}
On the other hand, the bond contribution within the BOCE method is described as
an analytical function of the bond order $P_{AB}$ between atoms $A$ and $B$,
as  obtained within the HF calculation. Therefore, to lowest order in $P_{AB}$,
one has \cite{Grassi:96,Grassi:04a}
\begin{equation}
E_c^{\mathrm{BOCE}} (AB) = a_{AB} P_{AB},
\label{eq:cBOCE}
\end{equation}
where $a_{AB}$ is a parameter which depends on the $A-B$ bond. Summarizing, the
theoretical molecular correlation energy within the BOCE method can be
expressed as
\begin{equation}
E_c^{\mathrm{BOCE}} = \sum_A E_c^{\exp} (A) + \sum_{\mbox{\scriptsize all $AB$}}
a_{AB} P_{AB} .
\label{eq:BOCE}
\end{equation}

The bonding parameters $a_{AB}$ in the second term of Eq.~(\ref{eq:BOCE}) can
be obtained from the experimental correlation energies of model molecules. In
particular, having $M$ different pairs of atoms in a molecule, each of them
contained $n_i$ times, we can rewrite the second term of Eq.~(\ref{eq:BOCE}) as
\begin{equation}
\sum_{\mbox{\scriptsize all $AB$}} a_{AB} P_{AB}
=
n_1 a_{AB} P_{AB} + n_2 a_{AC} P_{AC}  + \ldots + n_M a_{WZ} P_{WZ} ,
\end{equation}
where it has been assumed that the bond-order of each pair is approximately the
same for each pair of the same kind in the molecule. In this sense, the
parameter $a_{AB}$ is the binding correlation energy per bond $A-B$.

The procedure starts with the calculation of the parameter $a_{\mathrm{HH}}$
obtained from the experimental correlation energy of a H$_2$ molecule,
\begin{equation}
a_{\mathrm{HH}} = \frac{E_c^{\exp} (\mathrm{H}_2 ) - 2 E_c^{\exp}
(\mathrm{H})}{P_{\mathrm{HH}}} .
\end{equation}
In a similar way, the parameter $a_{\mathrm{CH}}$ is obtained from the
experimental correlation energy of the CH$_4$ molecule, and making use of the
value for $a_{\mathrm{HH}}$,
\begin{eqnarray}
a_{\mathrm{CH}} &=& \frac{E_c^{\exp} (\mathrm{CH}_4 ) - [ 4 E_c^{\exp}
(\mathrm{H}) + E_c^{\exp} (\mathrm{C} ) + 6 a_{\mathrm{HH}} P_{\mathrm{HH}} ]}{4
P_{\mathrm{CH}}} \nonumber\\
&=&
\frac{E_c^{\exp} (\mathrm{CH}_4 ) -6 a_{\mathrm{HH}} P_{\mathrm{HH}}}{4
P_{\mathrm{CH}}} .
\label{eq:aCH}
\end{eqnarray}
Following an analogous procedure, the parameters $a_{\mathrm{CC}}$,
$a_{\mathrm{CO}}$, $a_{\mathrm{OO}}$, and $a_{\mathrm{OH}}$ have been obtained
starting from the experimental correlation energies of C$_2$H$_6$,
(CH$_3$)$_2$O, H$_2$O$_2$, and H$_2$O, respectively (see
Table~\ref{tab:bondparams}).

In concluding this section, let us briefly recall the definition of the
thermodynamical quantities which will be calculated in the next section,
\emph{viz.} the molecular dissociation energy and the molecular heat of
formation.

\subsection{Definition of some relevant thermodynamic quantities}

The experimental molecular dissociation energy is defined as
\begin{equation}
D_0^{\exp} = \sum_A E_S (A) -E_S + E_{ZPE} ,
\end{equation}
where $E_S (A)$ is the Schr\"odinger energy of atom $A$ in the molecule, $E_S$
is the total Schr\"odinger energy of the molecule, and $E_{ZPE}$ is the 
vibrational Zero Point Energy, which is calculated from the values of the
molecular vibrational frequencies. In view of Eq.~(\ref{eq:identify}), within
the BOCE approximation one has for the molecular dissociation energy:
\begin{equation}
D_0^{\mathrm{BOCE}} = \sum_A E_S (A) -E^{\mathrm{BOCE}} + E_{ZPE} .
\label{eq:D0BOCE}
\end{equation}
Therefore, the difference between the experimental and calculated dissociation
energies can be expressed as
\begin{eqnarray}
D_0^{\exp} - D_0^{\mathrm{BOCE}} 
&=& E^{\mathrm{BOCE}} - E_S \nonumber\\
&=& E_c^{\mathrm{BOCE}} - E_c^{\exp} \nonumber\\
&=& \sum_{\mbox{\scriptsize all $AB$}} [
E_c^{\mathrm{BOCE}} (AB) - E_c^{\exp} (AB) ].
\label{eq:D0BOCEdiff}
\end{eqnarray}
In other words, only the difference between bonding BOCE and experimental
energies contributes to $D_0^{\exp} - D_0^{\mathrm{BOCE}}$.

Finally, following Cremer \emph{et al.} \cite{Cremer:82,Cremer:82a}, the
experimental molecular heat of formation is related to $E_S$ by
\begin{eqnarray}
\Delta H^{\exp} &=& (H_{\mbox{\scriptsize 298~K}} - H_{\mbox{\scriptsize 0~K}} ) -
\left( \sum_A E_S (A) -E_S + E_{ZPE} \right) \nonumber\\
&=& (H_{\mbox{\scriptsize 298~K}} - H_{\mbox{\scriptsize 0~K}} ) - D_0^{\exp} ,
\label{eq:Hexp}
\end{eqnarray}
and therefore the BOCE molecular heat of formation is related to
$E^{\mathrm{BOCE}}$ by
\begin{eqnarray}
\Delta H^{\mathrm{BOCE}} &=& (H_{\mbox{\scriptsize 298~K}} - H_{\mbox{\scriptsize 0~K}} ) -
\left( \sum_A E_S (A) -E^{\mathrm{BOCE}} + E_{ZPE} \right) \nonumber\\
&=& (H_{\mbox{\scriptsize 298~K}} - H_{\mbox{\scriptsize 0~K}} ) - D_0^{\mathrm{BOCE}} .
\label{eq:HBOCE}
\end{eqnarray}
In the above equations, $H_{\mbox{\scriptsize 298~K}} - H_{\mbox{\scriptsize
0~K}}$ is the variation of molecular enthalpy from 298~K to 0~K, and can be
evaluated according to standard thermodynamical formulas \cite{Nicolaides:96}.
Therefore, subtracting Eqs.~(\ref{eq:Hexp}) and (\ref{eq:HBOCE}), one obtains
\begin{equation}
\Delta H^{\exp} - \Delta H^{\mathrm{BOCE}} = D_0^{\mathrm{BOCE}} - D_0^{\exp} .
\label{eq:DeltaHBOCE}
\end{equation}

We recall that the above derivation holds in the BOCE approximation, where $E_S
= E^{\mathrm{BOCE}}$, for atoms. In general, for other methods, such as the G2
method, one has $E_S \neq E^{\mathrm{G2}}$, and Eqs.~(\ref{eq:D0BOCE}) and
(\ref{eq:D0BOCEdiff}) become:
\begin{equation}
D_0^{\mathrm{G2}} = \sum_A E^{\mathrm{G2}} (A) -E^{\mathrm{G2}} + E_{ZPE} ,
\end{equation}
and
\begin{equation}
D_0^{\exp} - D_0^{\mathrm{G2}} =
\sum_A [ E_S (A) - E^{\mathrm{G2}} (A) ] + (E_S - E^{\mathrm{G2}} ),
\label{eq:DeltaDG2}
\end{equation}
respectively, whereas Eq.~(\ref{eq:DeltaHBOCE}) holds also for the G2 method,
\emph{i.e.}
\begin{equation}
\Delta H^{\exp} - \Delta H^{\mathrm{G2}} = D_0^{\mathrm{G2}} - D_0^{\exp} .
\label{eq:DeltaHG2}
\end{equation}

\section{Results and discussion}

The procedures involved in all G$x$ calculations ($x=1,2,3$) are quite similar,
and have been described in the original works by Curtiss \emph{et al.}
\cite{Curtiss:91}. With respect to the calculation of the molecular heats of
formation, the results of the G$x$ methods show that, in some classes of
compounds, the G3 method is more accurate, while for other classes of compounds
the G2 or the G1 methods seem to be more accurate than the G3. We have chosen
to compare the molecular heats of formation calculated within the BOCE
approximation with those obtained with the G2 method.

In all calculations involving the BOCE method, we have used the basis sets 
3-21G, 6-311G, 6-31G$^{\ast\ast}$, 6-311G$^{\ast\ast}$, and
6-311G$++^{\ast\ast}$, and in the calculation of the bond-order matrix we have
employed L\"owdin's definition \cite{Loewdin:50} for all the latter basis sets.
Our results for the bond parameters $a_{AB}$ for the different chosen basis
sets are reported in Tab.~\ref{tab:bondparams}, while our results for the
bonding correlation energies, $E^{\mathrm{BOCE}} (AB)$ are reported in
Tab.~\ref{tab:bondenergies}, for all pairs $AB$ between H, C, and O. It may be
seen that the variations in the bond parameters from one basis set to another
are closely correlated with those in the bonding correlation energies for the
same bond and between the same basis sets. This is in agreement with the
procedure outlined in the previous section [see \emph{e.g.}
Eq.~(\ref{eq:aCH})], in view of the minor dependence of $a_{AB}$ on the
bond-order matrix element $P_{AB}$ than on the experimental molecular
correlation energy $E_c^{\exp}$.

\subsection{Molecular energies}

Tab.~\ref{tab:molecules} lists the set of 23 molecules, containing C, O, and H,
considered in this work. For these molecules, Tab.~\ref{tab:energies} reports
the experimental values of the Schr\"odinger energies, the total molecular
energy (including correlation) within the G2 model, $E^{\mathrm{G2}}$, and the
total molecular energy (again, including correlation) within the BOCE
approximation, $E^{\mathrm{BOCE}} = E_c^{\mathrm{BOCE}} + E_\HF$. The latter
depends on the particular basis set chosen for the HF calculation, as described
above.

It may be seen that the difference $|E^{\mathrm{G2}} - E_S |$ ranges between
$\approx 140$ and 535~mhartree, whereas $|E^{\mathrm{BOCE}} - E_S |$ is always
below $\approx25$~mhartree, for all the basis sets considered. Our first
conclusion is therefore that, with respect to the Schr\"odinger result, the
BOCE approximation is more accurate than the G2 method for the 23 molecules in
our data set. While it is tempting to assume that this conclusion will remain
true for the entire G2 data set, this is not decisive based solely on our
present study. In particular, from Tab.~\ref{tab:energies}, the best BOCE
results correspond to the 6-311G basis set. This is due to a competition of
various effects. First of all, the HF contribution to the total energy
$E^{\mathrm{BOCE}} = E_c^{\mathrm{BOCE}} + E_\HF$ clearly decreases on
increasing the size of the basis set, owing to the variational nature of the HF
calculation. Therefore, one invariably finds that $E_\HF$ decreases along the
series 3-21G~$\to$~6-311G$++^{\ast\ast}$. On the other hand, following
Eq.~(\ref{eq:Etheo}), and decomposing the correlation energy
$E_c^{\mathrm{BOCE}}$ into atomic and binding contributions, one finds that the
atomic (respectively, binding) contributions for the various basis sets are
minimum (respectively, maximum) for the 6-311G basis set. This is probably due
to the nature of the 6-311G basis set (and of the 6-3xx basis set family, in
general), usually employed to optimize the energies of the single atoms.

As a result of the compensation of these various contributions, the use of the
6-311G basis set within the BOCE method yields the best agreement with the
Schr\"odinger energy. This is especially important, in view of the dependence
of the dissociation energies $D_0$ on $E_S$.

In concluding this subsection on the molecular energies, let us discuss the
dependence of the bond order for a given bond on the basis set employed. The
average values of the bond order $P_{AB}$ for the bonds C--H, C--C, O--H, C--O
appearing in the molecules listed in Tab.~\ref{tab:molecules} have been
reported in Tab.~\ref{tab:bondorders} for the various basis sets.  (The bond
O--O has not been included, since there is only one molecule in
Tab.~\ref{tab:molecules} involving such bond.) One can see that, in bonds
involving H and C atoms, the bond order variation along the series of basis
sets considered here is below 4\%, whereas in bonds involving the oxygen atom,
the bond order variation ranges from $\approx 18$\% (C--O) to $\approx 27$\%
(O--H). This is in agreement with the study of Sannigrahi \cite{Sannigrahi:92},
where a more pronounced dependence of the bond-order and of the valency on the
basis set is found for all compounds containing electronegative centers (such
as O, F, Cl). For calculations concerning these compounds, Sannigrahi
\cite{Sannigrahi:92} therefore recommends the use of double-zeta basis sets,
and of polarization and diffuse functions in the case of highly ionic and/or
negatively charged species.

\subsection{Dissociation energy and molecular heat of formation}

In Tab.~\ref{tab:dissociations} we report the experimental and theoretical
molecular dissociation energies $D_0$ for the set of molecules listed in
Tab.~\ref{tab:molecules}. In particular, Tab.~\ref{tab:dissociations} compares
the theoretical values obtained within the G2 method (second column) and the
BOCE approximation, for the five basis sets considered above. Analogously,
Tab.~\ref{tab:enthalpies} reports the experimental and theoretical molecular
heats of formation $\Delta H$ for the same molecules, models, and basis sets.

In the case of the theoretical dissociation energy calculated within the G2
method, $D_0^{\mathrm{G2}}$, we find an average percentage difference with
respect to the experimental value of $\approx 0.2$~\% (corresponding to
$\approx 1.5$~kcal). For the same quantity calculated within the BOCE
approximation, we find an average percentage difference with respect to the
experimental value of $\approx 0.4 - 0.8$~\% (corresponding to $\approx 2.4 -
5.9$~kcal), depending on the basis set employed (Tab.~\ref{tab:dissociations}).
Analogously, for the molecular heats of formation (Tab.~\ref{tab:enthalpies}),
on the average we find $|\Delta H^{\mathrm{G2}} - \Delta H^{\exp} | \approx
8.2$~\% (corresponding to $\approx 1.5$~kcal), and $|\Delta H^{\mathrm{BOCE}} -
\Delta H^{\exp} | \approx 21.4-58.5$~\% (corresponding to $\approx
2.4-5.9$~kcal). (Such larger values of the average percentage errors are
justified by the smaller values of the experimental molecular heats of
formation, than the experimental molecular dissociation energies.) In both
cases, the best agreement within the BOCE approximation is obtained for the
6-311G basis set, as it was the case for the molecular energies.

Therefore, we may conclude that, while the G2 method yields in general more
accurate estimates of the molecular dissociation energies and the heats of
formation, the BOCE approximation affords theoretical estimates of the above
quantities of comparable accuracy, but now requiring much less computational
effort. Within the BOCE approximation, moreover, the best agreement with the
experimental results is obtained for the 6-311G basis set. As observed for the
molecular energies, this is a result of the competition of the various atomic
and binding contributions to the correlation energies (and therefore to the
other thermodynamical quantities considered in this work). In particular,
within the BOCE approximation, the binding contribution to the total
correlation energy is relatively larger than the atomic contribution, and it
can be therefore directly related to the bond-order value.

While it is agreed that the use of modest basis sets reduce the computing time
required to obtain chemical accuracy relative to the G$x$ methods, it is
appropriate, at this point, to expand on our reasoning as to why the 6-311G
basis set gives better results than the somewhat larger basis sets we have also
worked with. For such, still quite restricted, basis sets, there is rather
non-uniform `convergence'. We offer, as an explanation of such non-uniformity,
that the better results obtained with the 6-311G basis set rest on a somewhat
delicate interplay between diffuse $s$-like ($+$) and polarization $d$-like
($\ast$) functions on the one hand, and functions corresponding to valence
electrons on the other. As a consequence, in the larger basis sets considered
here, the total bond-order of each atom is overestimated with respect to the
6-311G basis set. The latter point about bond-order can be illustrated with
reference to Table~\ref{tab:bondorders}. If we neglect the 3-21G basis set
recorded there, then with just two exceptions (the C--C bond with the
6-311G$^{\ast\ast}$ basis set and the C--H bond with the 6-311G$++^{\ast\ast}$
basis set), the bond orders are smallest for the 6-311G set. Specifically then,
the advantage among these small basis sets lies with the 6-311G basis set in
that its diffuse and polarization functions increase, in the final bond-order
matrix, the total number of electrons in each atom contributing to the bond.

\begin{figure}[th]
\centering
\includegraphics[height=0.8\columnwidth,angle=-90]{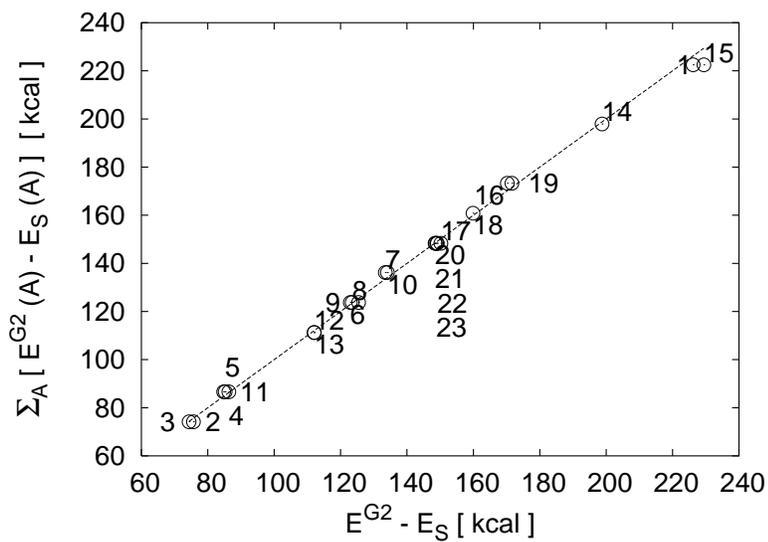}
\caption{Atomic \emph{vs} molecular contributions to the difference between
experimental and theoretical dissociation energy, Eq.~(\ref{eq:DeltaDG2}),
within the G2 method, for the 23 molecules listed in Tab.~\ref{tab:molecules}.
The dashed line is a guide to the eye.}
\label{fig:diffs}
\end{figure}

A finer analysis of the various terms contributing to the molecular
dissociation energies and heats of formation reveals that the very good
agreement of the G2 method with the experimental results is due to a partial
compensation of the two competing contributions for $D_0^{\exp} -
D_0^{\mathrm{G2}}$ and $\Delta H^{\exp} - \Delta H^{\mathrm{G2}}$ in the right
hand sides of Eqs.~(\ref{eq:DeltaDG2}) and (\ref{eq:DeltaHG2}), respectively.
Indeed, one finds $E^{\mathrm{G2}} -E_S \approx 74.356 - 229.413$~kcal, while 
$\sum_A [ E^{\mathrm{G2}} (A) -E_S (A) ] \approx 74.169 - 222.507$~kcal, for
the various molecules of Tab.~\ref{tab:molecules}, with differences between the
two quantities ranging between $-3.0$~kcal and $6.9$~kcal
(Fig.~\ref{fig:diffs}). Therefore, in some way, within the G2 method the errors
on the molecular part are compensated by the atomic part, \emph{i.e.} the
binding energy is very close to the Schr\"odinger value.

On the contrary, within the BOCE method [cf. Eqs.~(\ref{eq:D0BOCEdiff}) and
(\ref{eq:DeltaHBOCE})], the difference between experimental and calculated
molecular dissociation energies and heats of formation depends only on the
difference between the Schr\"odinger and the BOCE molecular energy, given the
assumption that the BOCE atomic energies are equal to the Schr\"odinger
counterparts, Eq.~(\ref{eq:identify}). Therefore, even though the BOCE
molecular energy is very close to the Schr\"odinger value, such a small
difference (in kcal) cannot be compensated by the atomic contribution.

\section{Conclusions and directions for future work}

We have compared and contrasted the G$x$ method (especially with $x=2$) and the
BOCE approximation for estimating \emph{(a)} molecular energies, \emph{(b)}
molecular dissociation energies, and \emph{(c)} molecular heats of formation
for some 20 molecules containing C, H, and O atoms. Although the G2 method
usually yields better agreement with the experimental results, we find that the
accuracy of the theoretical estimates of the above observable quantities within
the BOCE approximation is comparable to that of the G2 calculations, but now
with a remarkable saving in terms of computational complexity. With respect to
previous studies \cite{Grassi:96,Grassi:04a}, all BOCE parameters have been
calculated on single model molecules, rather than averaging over several
molecules. Therefore, each BOCE parameter $a_{AB}$ reflects the nature of the
particular bond $A-B$, irrespective of its chemical surroundings. In other
words, we find that the BOCE method is rather robust with respect to the
inclusion of further than binary correlation terms. Moreover, we have
extensively analyzed the dependence of the BOCE results on the basis set chosen
for the underlying HF calculation.

Owing to the much reduced computational effort required in a BOCE calculation,
we plan to apply the latter method to the evaluation of molecular properties of
larger molecules, where G$x$ methods are expected to require presently
prohibitive resources. It is also intended, in the light of the excellent
results for molecular dissociation energies, to study for such a set of larger
molecules the relation between the findings of the BOCE approximation and
Teller's theorem \cite{Teller:62}, which states that molecules do not bind in a
fully local density approximation (LDA), \emph{i.e.} including kinetic energy
density in LDA. Mucci and March \cite{Mucci:83} proposed to make a merit out of
Teller's theorem, by relating dissociation energy to molecular electron density
gradients \cite{Allan:85,Lee:86}. However, electron correlation in the
separated atoms seems important also in respect to heats of formation [see
Eqs.~(\ref{eq:DeltaDG2}) and (\ref{eq:DeltaHG2}) above], and this will also
require further exploration.

\subsection*{Acknowledgements}

NHM wishes to thank Professor V. E. Kravtsov for making possible his stay at
ICTP during 2005, where his contribution to this study was brought to fruition.

\bibliographystyle{molphys}
\bibliography{a,b,c,d,e,f,g,h,i,j,k,l,m,n,o,p,q,r,s,t,u,v,w,x,y,z,zzproceedings,Angilella,notes}

\begin{thebibliography}{10}

\bibitem{Curtiss:91}
Curtiss, L.~A., Raghavachari, K., Trucks, G.~W., and Pople, J.~A., 1991,
  \emph{J. Chem. Phys.}, \textbf{94}, 7221.

\bibitem{Moeller:34}
M\o{}ller, C. and Plesset, M.~S., 1934, \emph{Phys. Rev.}, \textbf{46}, 618.

\bibitem{note:computational}
For the molecules referred to in the present work (see Tab.~\ref{tab:molecules}
  below), G2 calculations required $\approx 200-800$~minutes of computation
  time on an Intel Pentium~IV (3.40~GHz), while BOCE calculations required
  $\approx 1-3$~minutes with the 6-311G basis set, and $\approx 8-12$~minutes
  with the more extended 6-311G$++^{\ast\ast}$ basis set. In both cases, the
  calculation of the BOCE correlation term required milliseconds. Concerning
  the storage memory required, Ref.~\cite{Curtiss:98} reports that 4.3~Gb were
  required for a G2 calculation on benzene.

\bibitem{Curtiss:98}
Curtiss, L.~A., Raghavachari, K., Redfern, P.~C., Rassolov, V., and Pople,
  J.~A., 1998, \emph{J. Chem. Phys.}, \textbf{109}, 7764.

\bibitem{Cremer:82}
Cremer, D., 1982, \emph{J. Comp. Chem.}, \textbf{3}, 154.

\bibitem{Cremer:82a}
Cremer, D., 1982, \emph{J. Comp. Chem.}, \textbf{3}, 165.

\bibitem{Grassi:96}
Grassi, A., Lombardo, G.~M., March, N.~H., and Pucci, R., 1996, \emph{Mol.
  Phys.}, \textbf{87}, 553.

\bibitem{Lewis:24}
Lewis, G.~N., 1924, \emph{J. Am. Chem. Soc.}, \textbf{46}, 2031.

\bibitem{Grassi:04a}
Grassi, A., Lombardo, G.~M., Forte, G., Angilella, G. G.~N., Pucci, R., and
  March, N.~H., 2004, \emph{accepted for publication in Mol. Phys.},
  \textbf{...}, ..., also available as preprint {\tt physics/0412066}.

\bibitem{Nicolaides:96}
Nicolaides, A., Rauk, A., Glukhovtsev, M.~N., and Radom, L., 1996, \emph{J.
  Phys. Chem.}, \textbf{100}, 17460.

\bibitem{Loewdin:50}
L\"owdin, P.~O., 1950, \emph{J. Chem. Phys.}, \textbf{18}, 365.

\bibitem{Sannigrahi:92}
Sannigrahi, A.~B., 1992, \emph{Adv. Quantum Chem.}, \textbf{23}, 301.

\bibitem{Teller:62}
Teller, E., 1962, \emph{Rev. Mod. Phys.}, \textbf{34}, 627.

\bibitem{Mucci:83}
Mucci, J.~F. and March, N.~H., 1983, \emph{J. Chem. Phys.}, \textbf{78}, 6187.

\bibitem{Allan:85}
Allan, N.~L., West, C.~G., Cooper, D.~L., Grout, P.~J., and March, N.~H., 1985,
  \emph{J. Chem. Phys.}, \textbf{83}, 4562.

\bibitem{Lee:86}
Lee, C. and Ghosh, S.~K., 1986, \emph{Phys. Rev. A}, \textbf{33}, 3506.

\end{thebibliography}

\newpage

\begin{table}[th]
\caption{Bond-order parameters $a_{AB}$ in Eq.~(\ref{eq:cBOCE}) (in mhartree)
for all possible bonds $A-B$ between C, O, and H. Their values depend on the HF
reference energy, which depends in turn on the particular choice of basis set.}
\label{tab:bondparams}
\begin{center}
\begin{small}
\begin{tabular}{|c|l|ccccc|}
\hline
$A-B$ & Model molecule & 3-21G & 6-311G & 6-31G$^{\ast\ast}$ & 6-311G$^{\ast\ast}$ &
6-311G$++^{\ast\ast}$ \\
\hline
H--H & H$_2$ 		& 43.86 &	46.00 &   39.55 &   41.55 &   41.55\\
C--H & CH$_4$ 		& 39.79 &	41.51 &   35.12 &   34.85 &   35.59\\
C--C & C$_2$H$_6$ 	& 39.77 &	39.60 &   31.70 &   30.68 &   30.06\\
O--H & H$_2$O 		& 88.43 &	84.63 &   60.54 &   51.20 &   51.71\\
C--O & (CH$_3$)$_2$O 	& 67.85 &	68.14 &   49.11 &   43.91 &   41.11\\
O--O & H$_2$O$_2$ 	& 83.33 &	89.52 &   72.04 &   69.92 &   66.62\\
\hline
\end{tabular}
\end{small}
\end{center}
\end{table}

\begin{sidewaystable}[th]
\caption{Bond correlation energies, $E_c (AB)$, in hartree, for all possible
bonds $A-B$ between C, O, and H. Their values depend on the HF reference
energy, which depends in turn on the particular choice of basis set.}
\label{tab:bondenergies}
\begin{center}
\begin{small}
\begin{tabular}{|c|l|r@{.}lr@{.}lr@{.}lr@{.}lr@{.}l|}
\hline
$A-B$ & Model molecule & \multicolumn{2}{c}{3-21G} & \multicolumn{2}{c}{6-311G}
& \multicolumn{2}{c}{6-31G$^{\ast\ast}$} &
\multicolumn{2}{c}{6-311G$^{\ast\ast}$} &
\multicolumn{2}{c|}{6-311G$++^{\ast\ast}$} \\
\hline
H--H & H$_2$ 		& $-1$ & 17442   &      $-1$ & 17442   &      $-1$ & 17442    &	$-1$ & 17442	 &$-0$ & 17442\\
C--H & CH$_4$ 		& $-40$ & 51353  &	    $-40$ & 51353  &      $-40$ & 51353   &	$-40$ & 51353	 &$-39$ & 51353\\
C--C & C$_2$H$_6$ 	& $-79$ & 82261  &	    $-79$ & 82261  &      $-79$ & 82261   &	$-79$ & 82261	 &$-78$ & 82261\\
O--H & H$_2$O 		& $-76$ & 43076  &	    $-76$ & 43076  &      $-76$ & 43076   &	$-76$ & 43076	 &$-75$ & 43076\\
C--O & (CH$_3$)$_2$O 	& $-151$ & 54948 &      $-151$ & 54948 &      $-151$ & 54948  &	$-151$ & 54948	 &$-150$ & 54948\\
O--O & H$_2$O$_2$ 	& $-155$ & 02013 &      $-155$ & 02013 &      $-155$ & 02013  &	$-155$ & 02013	 &$-154$ & 02013\\
\hline
\end{tabular}
\end{small}
\end{center}
\end{sidewaystable}

\begin{table}[th]
\caption{List of the 23 molecules, containing C, O, H, considered in this work.}
\label{tab:molecules}
\begin{center}
\begin{small}
\begin{tabular}{|r|l|l|}
\hline
1 &benzene	    & C$_6$H$_6$	   \\
2 &ethylene	    & C$_2$H$_4$	   \\
3 &acetylene	    & C$_2$H$_2$	   \\
4 &formaldehyde	    & HCHO		   \\
5 &methyl alcohol   & CH$_3$OH  	   \\
6 &ketene 	    & CH$_2$CO  	   \\
7 &carbon dioxide     & CO$_2$		   \\
8 &acetaldehyde	    & CH$_3$CHO 	   \\
9 &ethenol	    & CH$_2$CHOH	   \\
10&formic acid	    & HCOOH		   \\
11&carbon monoxide    & CO		   \\
12&cyclopropane	    & (CH$_2$)$_3$	   \\
13&1,2 propadiene 	    & CH$_2$CCH$_2$	   \\
14&furan 	    & C$_4$H$_4$O	   \\
15&cyclohexane	    & (CH$_2$)$_6$	   \\
16&glyoxal	    & HCOCOH		   \\
17&1,3 butadiene      & CH$_2$CHCHCH$_2$   \\  
18&acetone	    & CH$_3$COCH$_3$	   \\
19&acetic acid	    & CH$_3$COOH	   \\
20&propene	    & CH$_3$CHCH$_2$	   \\
21&2-butene (E)	    & $t$-CH$_3$CHCHCH$_3$ \\
22&2-butene (Z)	    & $c$-CH$_3$CHCHCH$_3$ \\
23&2-methyl 1-propene & C(CH$_3$)$_2$CH$_2$\\  
\hline
\end{tabular}
\end{small}
\end{center}
\end{table}

\begin{sidewaystable}[th]
\caption{Total molecular energies (in hartree), $E^{\mathrm{BOCE}} =
E_c^{\mathrm{BOCE}} + E_\HF$, including correlation, for the 23 molecules
listed in Tab.~\ref{tab:molecules}. The various columns refer to the different
basis sets being considered in the HF calculation. The second and third columns
are the total molecular Schr\"odinger energy, $E_S$, and $E^{\mathrm{G2}}$,
respectively.}
\label{tab:energies}
\begin{center}
\begin{small}
\begin{tabular}{|c|r@{.}l|r@{.}l|r@{.}lr@{.}lr@{.}lr@{.}lr@{.}l|}
\hline
 & 
 \multicolumn{2}{|c}{$E_S$} &
 \multicolumn{2}{|c|}{$E^{\mathrm{G2}}$} &
 \multicolumn{10}{c|}{$E^{\mathrm{BOCE}}$} \\
 \cline{6-15}
 & \multicolumn{2}{c|}{} & \multicolumn{2}{c|}{} &
 \multicolumn{2}{c}{3-21G} &
 \multicolumn{2}{c}{6-311G} &
 \multicolumn{2}{c}{6-31G$^{\ast\ast}$} &
 \multicolumn{2}{c}{6-311G$^{\ast\ast}$} &
 \multicolumn{2}{c|}{6-311G$++^{\ast\ast}$} \\
 \hline
1 &$-232$ & 24219 &$-231$ & 78054 &$-232$ & 23115 &$-232$ & 24129 &$-232$ & 22161 &$-232$ & 22495 &$-232$ & 21644\\	
2 &$-78$ & 58567  &$-78$ & 41594  &$-78$ & 58344  &$-78$ & 58651  &$-78$ & 58084  &$-78$ & 58302  &$-78$ & 58157 \\	
3 &$-77$ & 33430  &$-77$ & 18574  &$-77$ & 33024  &$-77$ & 33251  &$-77$ & 31958  &$-77$ & 32522  &$-77$ & 32114 \\	
4 &$-114$ & 50344 &$-114$ & 33892 &$-114$ & 49660 &$-114$ & 50112 &$-114$ & 50278 &$-114$ & 50176 &$-114$ & 49490\\	
5 &$-115$ & 72137 &$-115$ & 53489 &$-115$ & 72411 &$-115$ & 72416 &$-115$ & 72340 &$-115$ & 72350 &$-115$ & 72202\\	
6 &$-152$ & 6	  &$-152$ & 36911 &$-152$ & 60331 &$-152$ & 59168 &$-152$ & 59566 &$-152$ & 59897 &$-152$ & 58538\\
7 &$-188$ & 58504 &$-188$ & 36132 &$-188$ & 58684 &$-188$ & 57255 &$-188$ & 60229 &$-188$ & 60371 &$-188$ & 58161\\	
8 &$-153$ & 82924 &$-153$ & 57685 &$-153$ & 82178 &$-153$ & 82718 &$-153$ & 82682 &$-153$ & 82746 &$-153$ & 82040\\	
9 &$-153$ & 81314 &$-153$ & 55770 &$-153$ & 81266 &$-153$ & 81354 &$-153$ & 80838 &$-153$ & 81022 &$-153$ & 80530\\	
10&$-189$ & 76370 &$-189$ & 51649 &$-189$ & 77233 &$-189$ & 76925 &$-189$ & 77838 &$-189$ & 77720 &$-189$ & 76468\\	
11&$-113$ & 31751 &$-113$ & 17750 &$-113$ & 31573 &$-113$ & 31644 &$-113$ & 32008 &$-113$ & 32148 &$-113$ & 31558\\	
12&$-117$ & 89065 &$-117$ & 63115 &$-117$ & 87292 &$-117$ & 88093 &$-117$ & 88280 &$-117$ & 88375 &$-117$ & 88177\\	
13&$-116$ & 65212 &$-116$ & 41784 &$-116$ & 64590 &$-116$ & 64947 &$-116$ & 63963 &$-116$ & 64355 &$-116$ & 63889\\	
14&$-230$ & 01980 &$-229$ & 63261 &$-230$ & 01332 &$-230$ & 01476 &$-230$ & 00899 &$-230$ & 00925 &$-229$ & 99859\\	
15&$-235$ & 87496 &$-235$ & 33937 &$-235$ & 85806 &$-235$ & 86391 &$-235$ & 86104 &$-235$ & 86280 &$-235$ & 86424\\	
16&$-227$ & 81908 &$-227$ & 51024 &$-227$ & 81254 &$-227$ & 82108 &$-227$ & 82216 &$-227$ & 82183 &$-227$ & 80646\\	
17&$-155$ & 98889 &$-155$ & 66427 &$-155$ & 97929 &$-155$ & 98644 &$-155$ & 97466 &$-155$ & 97879 &$-155$ & 97500\\	
18&$-193$ & 15209 &$-192$ & 81368 &$-193$ & 14508 &$-193$ & 15078 &$-193$ & 14894 &$-193$ & 14943 &$-193$ & 14242\\	
19&$-229$ & 08873 &$-228$ & 75391 &$-229$ & 09387 &$-229$ & 09385 &$-229$ & 09930 &$-229$ & 09921 &$-229$ & 08686\\	
20&$-117$ & 90274 &$-117$ & 64509 &$-117$ & 89730 &$-117$ & 90190 &$-117$ & 89541 &$-117$ & 89816 &$-117$ & 89658\\	
21&$-157$ & 21942 &$-156$ & 87484 &$-157$ & 20633 &$-157$ & 21660 &$-157$ & 21007 &$-157$ & 21245 &$-157$ & 18493\\	
22&$-157$ & 21848 &$-156$ & 87266 &$-157$ & 20984 &$-157$ & 21412 &$-157$ & 20760 &$-157$ & 21001 &$-157$ & 20536\\	
23&$-157$ & 22188 &$-156$ & 87670 &$-157$ & 21280 &$-157$ & 21716 &$-157$ & 21044 &$-157$ & 21311 &$-157$ & 18634\\	
\hline
\end{tabular}
\end{small}
\end{center}
\end{sidewaystable}

\begin{table}[th]
\caption{Dependence of the bond-order $P_{AB}$ for a given bond $A-B$ on the
basis set employed. Reported are the average bond-orders $P_{AB}$ for the bonds
more frequently appearing in the molecules listed in Tab.~\ref{tab:molecules}.}
\label{tab:bondorders}
\begin{center}
\begin{small}
\begin{tabular}{|c|ccccc|}
\hline
 & 3-21G & 6-311G & 6-31G$^{\ast\ast}$ & 6-311G$^{\ast\ast}$ &
 6-311G$++^{\ast\ast}$ \\
\hline
C--H &    0.94649 & 0.93456 & 0.94714 & 0.95081 & 0.91798\\
C--C &    1.05140 & 1.06971 & 1.07062 & 1.05695 & 1.09041\\
O--H &    0.92589 & 0.91894 & 1.00533 & 1.16700 & 1.12736\\
C--O &    1.32735 & 1.15963 & 1.24610 & 1.32735 & 1.37367\\
\hline
\end{tabular}
\end{small}
\end{center}
\end{table}

\begin{table}[th]
\caption{Experimental and calculated dissociation energies $D_0$ (in kcal) for
the 23 molecules listed in Tab.~\ref{tab:molecules}. Boldface values refer to
the method yielding the best agreement with the experimental result, for each
molecule. The two last lines refer to the average absolute difference and the
average percentage difference, respectively, between the experimental and
calculated $D_0$.}
\label{tab:dissociations}
\begin{center}
\begin{small}
\begin{tabular}{|c|r@{.}l|r@{.}l|r@{.}lr@{.}lr@{.}lr@{.}lr@{.}l|}
\hline
 & 
 \multicolumn{2}{|c}{$D_0^{\exp}$} &
 \multicolumn{2}{|c|}{$D_0^{\mathrm{G2}}$} &
 \multicolumn{10}{c|}{$D_0^{\mathrm{BOCE}}$} \\
 \cline{6-15}
 & \multicolumn{2}{c|}{} & \multicolumn{2}{c|}{} &
 \multicolumn{2}{c}{3-21G} &
 \multicolumn{2}{c}{6-311G} &
 \multicolumn{2}{c}{6-31G$^{\ast\ast}$} &
 \multicolumn{2}{c}{6-311G$^{\ast\ast}$} &
 \multicolumn{2}{c|}{6-311G$++^{\ast\ast}$} \\
 \hline
1 &1305 & 5  &1301 & 8  &1298 & 6  &{\bf 1305} & {\bf 0}  &1292 & 6  &1294 & 7  &1289 & 4\\
2 &531 & 9   &{\bf 531} & {\bf 7}   &530 & 5   &532 & 4   &528 & 8   &530 & 2	&529 & 3 \\
3 &388 & 8   &387 & 2   &386 & 2   &{\bf 387} & {\bf 6}   &379 & 5   &383 & 1	&380 & 5 \\
4 &359 & 0   &{\bf 359} & {\bf 3}   &354 & 7   &357 & 6   &358 & 6   &358 & 0	&353 & 7 \\
5 &480 & 9   &482 & 3   &482 & 6   &482 & 6   &482 & 1   &482 & 2	&{\bf 481} & {\bf 3} \\
6 &512 & 8   &{\bf 513} & {\bf 6}   &516 & 6   &509 & 3   &511 & 8   &513 & 8	&505 & 3 \\
7 &381 & 9   &384 & 6   &{\bf 383} & {\bf 1}   &374 & 1   &392 & 8   &393 & 7	&379 & 8 \\
8 &643 & 7   &{\bf 643} & {\bf 9}   &639 & 0   &642 & 4   &642 & 2   &642 & 6	&638 & 2 \\
9 &633 & 6   &631 & 9   &{\bf 633} & {\bf 3}   &{\bf 633} & {\bf 9}   &630 & 6   &631 & 8	&628 & 7 \\
10&480 & 0   &482 & 0   &485 & 4   &483 & 4   &489 & 2   &488 & 4	&{\bf 480} & {\bf 6} \\
11&256 & 2   &258 & 0   &255 & 1   &{\bf 255} & {\bf 5}   &257 & 8   &258 & 7	&255 & 0 \\
12&802 & 9   &{\bf 802} & {\bf 1}   &791 & 7   &796 & 8   &797 & 9   &798 & 5	&797 & 3 \\
13&669 & 1   &668 & 2   &665 & 1   &{\bf 667} & {\bf 4}   &661 & 2   &663 & 7	&660 & 7 \\
14&950 & 5   &{\bf 949} & {\bf 6}   &946 & 4   &947 & 3   &943 & 7   &943 & 9	&937 & 2 \\
15&1659 & 4  &1652 & 5  &1648 & 8  &1652 & 5  &1650 & 7  &1651 & 8  &{\bf 1652}& {\bf 7}\\
16&610 & 4   &613 & 4   &606 & 3   &{\bf 611} & {\bf 7}   &612 & 4   &612 & 2	&602 & 5 \\
17&960 & 1   &958 & 2   &954 & 1   &{\bf 958} & {\bf 6}   &951 & 2   &953 & 8	&951 & 4 \\
18&927 & 0   &{\bf 927} & {\bf 9}   &922 & 6   &926 & 2   &925 & 0   &925 & 3	&920 & 9 \\
19&764 & 7   &766 & 3   &767 & 9   &767 & 9   &771 & 3   &771 & 3	&{\bf 763} & {\bf 5} \\
20&811 & 2   &{\bf 810} & {\bf 8}   &807 & 8   &810 & 6   &806 & 6   &808 & 3	&807 & 3 \\
21&1090 & 5  &{\bf 1090} & {\bf 3}  &1082 & 3  &1088 & 7  &1084 & 9  &1086 & 2  &1084 & 9\\
22&1089 & 8  &{\bf 1089} & {\bf 0}  &1084 & 4  &1087 & 0  &1083 & 2  &1084 & 7  &1083 & 2\\
23&1092 & 2  &{\bf 1091} & {\bf 5}  &1086 & 5  &1089 & 2  &1085 & 1  &1086 & 7  &1085 & 9\\
\hline
\hline
\multicolumn{3}{|l|}{Avg. $|\Delta D_0 |$} &
1 & 5 & 4 & 5 & 2 & 4 & 5 & 5 & 4 & 6 & 5 & 9 \\
\multicolumn{3}{|l|}{Avg. $|\Delta D_0 |$ (\%)} &
0 & 2 & 0 & 6 & 0 & 4 & 0 & 8 & 0 & 7 & 0 & 8 \\
\hline
\end{tabular}
\end{small}
\end{center}
\end{table}

\begin{sidewaystable}[th]
\caption{Experimental and calculated molecular heats of formation at 298~K,
$\Delta H$ (in kcal), for the 23 molecules listed in Tab.~\ref{tab:molecules}.
Boldface values refer to the method yielding the best agreement with the
experimental result, for each molecule. The two last lines refer to the average
absolute difference and the average percentage difference, respectively,
between the experimental and calculated $\Delta H$.}
\label{tab:enthalpies}
\begin{center}
\begin{small}
\begin{tabular}{|c|r@{.}l|r@{.}l|r@{.}lr@{.}lr@{.}lr@{.}lr@{.}l|}
\hline
 & 
 \multicolumn{2}{|c}{$\Delta H^{\exp}$} &
 \multicolumn{2}{|c|}{$\Delta H^{\mathrm{G2}}$} &
 \multicolumn{10}{c|}{$\Delta H^{\mathrm{BOCE}}$} \\
 \cline{6-15}
 & \multicolumn{2}{c|}{} & \multicolumn{2}{c|}{} &
 \multicolumn{2}{c}{3-21G} &
 \multicolumn{2}{c}{6-311G} &
 \multicolumn{2}{c}{6-31G$^{\ast\ast}$} &
 \multicolumn{2}{c}{6-311G$^{\ast\ast}$} &
 \multicolumn{2}{c|}{6-311G$++^{\ast\ast}$} \\
 \hline
1 &19 & 8    &23 & 6    &26 & 8    &{\bf 20} & {\bf 4}    &32 & 7    &30 & 6	&36 & 0	\\ 
2 &12 & 5    &{\bf 12} & {\bf 8}    &13 & 9    &12 & 0    &15 & 6    &14 & 2	&15 & 1	\\ 
3 &54 & 2    &55 & 8    &56 & 7    &{\bf 55} & {\bf 3}    &63 & 4    &59 & 9	&62 & 5	\\ 
4 &$-27$ & 7   &{$\mathbf{-28}$} & {\bf 0}	&$-23$ & 4   &$-26$ & 2   &$-27$ & 3   &$-26$ & 7  &$-22$ & 3  \\ 
5 &$-48$ & 0   &$-49$ & 4	&$-49$ & 8   &$-49$ & 8   &$-49$ & 3   &$-49$ &4 &{$\mathbf{-48}$} & {\bf 4}  \\ 
6 &$-11$ & 5   &{$\mathbf{-12}$} & {\bf 2}	&$-15$ & 2   &$-7$ & 9    &$-10$ & 4   &$-12$ & 5  &$-4$ & 0   \\ 
7 &$-94$ & 1   &$-95$ & 4	&{$\mathbf{-95}$} & {\bf 2}   &$-86$ & 2   &$-104$ & 9  &$-105$ & 8 &$-91$ & 9  \\ 
8 &$-40$ & 8   &{$\mathbf{-41}$} & {\bf 0}	&$-36$ & 1   &$-39$ & 5   &$-39$ & 3   &$-39$ & 7  &$-35$ & 3  \\ 
9 &$-30$ & 6   &$-29$ & 0	&$-30$ & 3   &{$\mathbf{-30}$} & {\bf 8}   &$-27$ & 6   &$-28$ & 8  &$-25$ & 7  \\ 
10&$-90$ & 5   &$-92$ & 6	&$-95$ & 9   &$-94$ & 0   &$-99$ & 7   &$-99$ &0 &{$\mathbf{-91}$} & {\bf 1}  \\ 
11&$-26$ & 4   &$-28$ & 2	&$-25$ & 3   &{$\mathbf{-25}$} & {\bf 7}   &$-28$ & 0   &$-28$ & 9  &$-25$ & 2  \\ 
12&12 & 7    &{\bf 13} & {\bf 5}    &23 & 9    &18 & 8    &17 & 7    &17 & 1	&18 & 3	\\ 
13&45 & 5    &46 & 4    &49 & 4    &{\bf 47} & {\bf 2}    &53 & 4    &50 & 9	&53 & 8	\\ 
14&$-8$ & 3    &{$\mathbf{-7}$} & {\bf 4}    &$-4$ & 2   &$-5$ & 1    &$-1$ & 5   &$-1$ & 7	 &5 & 0	 \\ 
15&$-29$ & 4   &$-22$ & 5   &$-18$ & 8  &$-22$ & 5   &$-20$ & 7  &$-21$ & 8 &{$\mathbf{-22}$} & {\bf 7}  \\ 
16&$-50$ & 7   &$-53$ & 6   &$-46$ & 6  &{$\mathbf{-51}$} & {\bf 9}   &$-52$ & 6  &$-52$ & 4   &$-42$ & 7  \\ 
17&26 & 0    &28 & 0    &32 & 0    &{\bf 27} & {\bf 5}    &34 & 9    &32 & 3	&34 & 7	\\ 
18&$-52$ & 2   &{$\mathbf{-53}$} & {\bf 0}   &$-47$ & 8   &$-51$ & 4   &$-50$ & 2   &$-50$ & 6 &$-46$ & 2  \\ 
19&$-103$ & 44 &$-105$ & 1  &$-106$ & 7  &$-106$ & 7  &$-110$ & 1  &$-110$ & 0&{$\mathbf{-102}$} & {\bf 3} \\ 
20&4 & 9     &{\bf 5} & {\bf 3}     &8 & 3     &5 & 4     &9 & 5     &7 & 8	&8 & 7	\\ 
21&$-2$ & 6    &{$\mathbf{-2}$} & {\bf 4}    &1 & 8   &$-0$ & 8    &3 & 0     &1 & 7	&3 & 0    \\ 
22&$-1$ & 8    &{$\mathbf{-1}$} & {\bf 0}    &3 & 6   &0 & 9   &4 & 8	 &3 & 4     &4 & 7    \\ 
23&$-4$ & 3    &{$\mathbf{-5}$} & {\bf 0}    &1 & 4   &$-1$ & 2    &2 & 9     &1 & 2	&2 & 1    \\ 
\hline
\hline
\multicolumn{3}{|l|}{Avg. $|\Delta H^{\mathrm{theo}} - \Delta H^{\exp}|$} &
1 & 5 & 4 & 3 & 2 & 4 & 5 & 5 & 4 & 6 & 5 & 9 \\
\multicolumn{3}{|l|}{Avg. $|\Delta H^{\mathrm{theo}} - \Delta H^{\exp}|$ (\%)} &
8 & 2 & 44 & 3 & 21 & 4 & 52 & 9 & 41 & 8 & 58 & 5 \\
\hline
\end{tabular}
\end{small}
\end{center}
\end{sidewaystable}

\clearpage

\end{document}